\documentclass[11pt]{article}

\usepackage{amsmath}
\usepackage{amssymb}
\usepackage{times}
\usepackage{dsfont}
\usepackage{theorem}
\usepackage{subfigure}

\def\=d{\,{\buildrel\rm def\over =}\,}

\newcommand{\be}{\begin{equation}}
\newcommand{\ee}{\end{equation}}
\newcommand{\bd}{\begin{displaymath}}
\newcommand{\ed}{\end{displaymath}}
\newcommand{\ad }{a^{\dagger}}
\newcommand{\bp }{b^{\dagger}}
\newcommand{\at }{{ \tilde {a} } }
\newcommand{\att }{{ \tilde{ a} }^{\dagger}}
\newcommand{\bt }{{ \tilde{ b} } }
\newcommand{\btt }{{ \tilde{ b} }^{\dagger}}
\newcommand{\ai }{ \tilde{ a}_i }
\newcommand{\aj }{ \tilde{ a}_j }
\newcommand{\ait }{ \tilde{ a}_i^{\dagger}}
\newcommand{\ajt }{ \tilde{ a}_j^{\dagger}}
\newcommand{\ec }{\eta^{\dagger}}

\newcommand{\eb }{\eta_b^{\dagger}}
\newcommand{\qi}{ q^{-1}}
\newcommand{\la }{\Lambda}

\title{How to commute}
\author{Andreas Walter Aste$^{a,b}$\\
Won Sang Chung$^{*}$\\
$\quad$\\
$^{a}$\emph{Department of Physics, University of Basel, 4056 Basel, Switzerland}\\
$^{b}$\emph{Paul Scherrer Institute, 5232 Villigen PSI, Switzerland}\\
$\quad$\\
$^{*}$\emph{Department of Physics and Research Institute of Natural Science,}\\
\emph{College of Natural Science,
Gyeongsang National University, Jinju 660-701, Korea}
}
\date{January 14, 2014}

\begin{document}
\maketitle

\begin{abstract}
\noindent A simple exposition of the rarely discussed fact that a set of free boson fields
describing different, i.e. kinematically different particle types can be quantized with
mutual anticommutation relations is given by the explicit construction of the Klein transformations
changing anticommutation relations into commutation relations. The q-analog of the presented
results is also treated. The analogous situation for two independent free fermion fields with mutual
commutation or anticommutation relations is briefly investigated.
\\
\vskip 0.1 cm \noindent {\bf Physics and Astronomy Classification Scheme (2010).} 11.10.-z - Field theory;
11.30.-j Symmetry and conservation laws
\end{abstract}

\section{Introduction}
All hitherto existing experimental evidence indicates that physical systems with one type of
integer spin particles solely obey the laws of Bose-Einstein statistics, whereas systems
with one type of half-odd integer spin particles respect Fermi-Dirac statistics.
The natural way to arrive at Bose-Einstein or Fermi-Dirac statistics is to describe
the particles by the help of quantum fields which commute or anticommute for space-like separations,
repectively. When one turns from the commutation relations for a given field to those between
different fields in the sense that the fields cannot be mapped by space-time transformations
onto each other, the situation becomes more complicated. One observes that `abnormal'
commutation relations in theories in which, e.g., two different integer spin fields anticommute, may arise,
but such theories possess special symmetries which allow to link them
to the case with regular commutation relations. In this paper, it is shown how this link can easily
be constructed from simple algebraic considerations for systems with a finite number
of degrees of freedom which can be generalized in a straightforward manner
to the case of inifinitely many degrees of freedom.\\

\noindent Starting from the well-known commutation relations of the rising and lowering operators of
two independent, i.e. non-interacting bosonic harmonic oscillators  
\begin{displaymath}
[a,a^\dagger]=1 \,  , \quad [b,b^\dagger]=1 \, , \label{bosonic_cr}
\end{displaymath}
where $a$ and $b$ annihilate the ground state
\begin{equation}
a| 0 \rangle = b | 0 \rangle = 0 \, ,
\end{equation}
one may impose `abnormal' mutual anticommutation relations given by
\begin{equation}
\{ a,b \} = ab+ba = \{ a,b^\dagger \} = 0 \, , \label{acr}
\end{equation}
consequently leading to the Hermitian conjugate relations
\begin{equation}
\{a^\dagger, b^\dagger \} = \{ a,b \}^\dagger =
\{ a, b^\dagger \}^\dagger = \{ a^\dagger , b \} = 0 \, .
\end{equation}
The commutation relations given by eq. (\ref{bosonic_cr}) fix the physical nature
of the phonons as bosons, since the creation operators $a^\dagger$ and $b^\dagger$
can create an arbitrary number of phonons, contrary to the fermionic case where the corresponding
anticommutation relations would imply the Pauli principle  $a^{\dagger \, 2}=b^{\dagger \, 2}=0$.
In the bosonic case, a state normalized to one containing $\tilde{n}$ phonons of type $a$ would be given,
e.g., by
\begin{equation}
| \tilde{n} \rangle = \frac{1}{\sqrt{\tilde{n}!}} a^{\dagger \, \tilde{n}} |0 \rangle \, .
\end{equation}

\noindent The anticommutation relations in eq. (\ref{acr}) can be converted into commutation relations
by mapping the algebra of $b$-type operators only
according to
\begin{equation}
a \mapsto \tilde{a}=a \, , \quad
b \mapsto \tilde{b}=\eta b \, , \quad b^\dagger \mapsto \tilde{b}^\dagger =
b^\dagger \eta = -\eta b^\dagger \, , \label{mapping_1}
\end{equation}
where the involutive, unitary and Hermitian phonon (or particle) number parity operator $\eta$
is defined via the phonon number operator
\begin{equation}
n=a^\dagger a + b^\dagger b
\end{equation}
by
\begin{equation}
\eta = (-1)^n = e^{i \pi n} = e^{-i \pi n} = \eta^{-1} = \eta^\dagger \, .
\end{equation}
$\eta$ anticommutes with $a$, $a^\dagger$, $b$, and $b^\dagger$, since the creation and annihilation operators
change the particle number by one.
Hence, the commutation relations in the $b$-sector are preserved, since $[\tilde{b},\tilde{b}]=0$ is trivially
fulfilled and from $\eta^2=1$ follows
\begin{equation}
[\tilde{b},\tilde{b}^\dagger] =[\eta b, b^\dagger \eta ] =
\eta b b^\dagger \eta - b^\dagger \eta^2 b=
[b, b^\dagger]= 1 \, ,
\end{equation}
however one now has mutual commutativity
\begin{displaymath}
[a,\tilde{b}] = a \eta b - \eta b a = -\eta a b - \eta b a = -\eta \{ a , b \} = 0 \, ,
\end{displaymath}
\begin{displaymath}
[a , \tilde{b}^\dagger] = a b^\dagger \eta - b^\dagger \eta a = \{ a, b^\dagger \} \eta = 0 \, ,
\end{displaymath}
i.e. the anticommutation relations between the $a$- and $b$-operators go over into commutation
relations by a change of phase conventions without changing the physical content of the theory.
Note, however, that the transformation according to eq. (\ref{mapping_1}) does not correspond to a
unitary transformation of the operators and the underlying Hilbert space of phonon states which would preserve
the commutation rules.\\

\noindent The way to achieve a situation where all operators fulfill standard commutation rules is,
of course, not unique.
E.g., introducing particle number and particle number parity operators for different phonon types
\begin{equation}
n_a=a^\dagger a \, , \quad n_b= b^\dagger b \, ,
\end{equation}
\begin{equation}
\eta_a = (-1)^{n_a} = \eta_a^{-1} = \eta_a^\dagger \, , \quad  \eta_b = (-1)^{n_b} = \eta_b^{-1} = \eta_b^\dagger 
\end{equation}
and new operators
\begin{equation}
a \mapsto \tilde{a}=\eta_b a \, , \quad a^\dagger \mapsto \tilde{a}^\dagger =
a^\dagger \eta_b = \eta_b a^\dagger \, , \quad
b \mapsto \tilde{b}=\eta_b b \, , \quad b^\dagger \mapsto \tilde{b}^\dagger =
b^\dagger \eta_b = -\eta_b b^\dagger \, , \label{mapping_2}
\end{equation}
also does the job. A further valid redefinition is given by
\begin{equation}
a \mapsto \tilde{a}= a \, , \quad
b \mapsto \tilde{b}=\eta_a b \, , \quad b^\dagger \mapsto \tilde{b}^\dagger =
b^\dagger \eta_a = \eta_a b^\dagger \, . \label{mapping_2}
\end{equation}
So-called Klein transformations as presented above have been introduced for the first time by Oskar Klein
\cite{Klein}. The abstract work on a quantum field theoretical level in connection with the
spin-statistics theorem given much later by Huzihiro Araki \cite{Araki} was the basis for a short
discussion given by Ray Streater and Arthur Wightman in their famous book on PCT, spin and statistics, and
all that \cite{PCT}.

\section{Several degrees of freedom}
In the case where $m$ different phonon types are created by operators $a_1^\dagger, \ldots a_m^\dagger$
with
\begin{equation}
[a_i,a_i^\dagger]=1 \, , \quad \{a_i,a_j \}= \{ a_i ,  a_j^\dagger \} = 0 \quad
\mbox{for} \, \, i \neq j \, ,
\end{equation}
the mutual anticommutation relations can be successively transformed into commutation relations by
the following sequence of transformations
\begin{displaymath}
a_1 \mapsto \tilde{a}_1=a_1 \, ,
\end{displaymath}
\begin{displaymath}
a_2 \mapsto \tilde{a}_2= \eta_1 a_2 \, ,
\end{displaymath}
\begin{displaymath}
a_3 \mapsto \tilde{a}_2= \eta_1 \eta_2 a_3 \, ,
\end{displaymath}
\begin{displaymath}
\ldots
\end{displaymath}
\begin{displaymath}
a_i \mapsto \tilde{a}_i= \eta_1 \cdot \ldots \cdot \eta_{i-1} a_i \, ,
\end{displaymath}
\begin{displaymath}
\ldots
\end{displaymath}
\begin{equation}
a_m \mapsto \tilde{a}_m= \eta_1 \cdot \ldots \cdot \eta_{m-1} a_m \, ,
\end{equation}
where
\begin{equation}
\eta_i= (-1)^{a_i^\dagger a_i} \, .
\end{equation}
This explicit construction shows that there is always the possibility to successively
remove all abnormal anticommutators from a theory describing different bosons only.
An analogous statement holds for the general case involving different different bosonic and different fermionic
particles. For the even more general quantum field theoretical case where one has infinitely many degrees of freedom,
the (normal) abnormal case of two different Fermi fields with vanishing (anti-)commutators is briefly discussed in the
last section. But before the situation discussed above shall be reanalyzed from a q-deformed point of view.

\section{How to q-commute}
Considering again the commutation relations of the raising and lowering operators of two independent bosonic 
harmonic oscillators
\be
[a, \ad ] = 1 \, , \quad [ b, \bp ] =1 \, , \quad [a,b] = [ a, \bp ] = [ \bp , \ad ] = [ b, \ad ] =0 \, ,
\ee
where $ a $ and $b$ annihilate the ground state  $a|0\rangle = b |0\rangle =0$, one may introduce the map 
\be
\at = a , \quad \bt = \eta b , ~~ \att = \ad, \quad \btt = \bp \ec = \bp \eta^{-1} = q \eta^{-1} \bp
\label{qmap}
\ee
where $\eta$ is a unitary q-parity operator defined with $\theta$ real as
\be
\eta = q^N =  e^{ i \theta N} \,  , \quad \ec = e^{ - i \theta N} = \eta^{-1} \, ,
\label{qmap2}
\ee
and 
\be
N =N_a + N_b =  \ad a + \bp b \, .
\ee
Introducing the following commutation relations
\bd
\eta a = \qi a \eta \,  , \quad \eta \ad = q \ad \eta \, , 
\ed
\be
\eta b = \qi b \eta \,  , \quad \eta \bp = q \bp \eta \, , 
\ee
leads to the algebra
\bd
[ \at , \att ] =1 \,  , \quad [ \bt, \btt ] =1 \, ,
\ed
\bd
[ \at, \bt]_q = \at \bt - q \bt \at =0 \, ,
\ed
\bd
[ \at , \btt ]_{\qi} = \at \btt - \qi \btt \at =0 \, ,
\ed
\be
[ \tilde{N}_a , \tilde{N}_b ] =0 \, ,
\ee
where
\be
\tilde{N}_a = \att \at , \quad \tilde{N}_b = \btt \bt \, .
\ee
Replacing  $ \eta = q^{N} $ with $ \eta_a = q^{N_a } $ in eq. (\ref{qmap2}) gives the same result.
Indeed, the map given in eq.  (\ref{qmap}) transforms commuting modes into q-commuting modes without changing
the boson algebra for each mode. For $q = -1 $, the map eq. (\ref{qmap}) leads to the case which has been introduced
for the first time by Klein \cite{Klein}, and studied in further detail by
Araki \cite{Araki}. \\

\noindent One may also define another mapping like
\be
\at = \eta_b a \, , \quad \bt = \eta_b  b \, , ~~ \att = \ad \eb \, , \quad \btt = \bp \eb \, ,
\ee
where $\eta_b = q^{N_b }$. Then we have the following commutation relations
\bd
[ \at \, , \att ] =1 , \quad [ \bt \, , \btt ] =1 \, ,
\ed
\bd
[ \at \, , \bt]_{\qi}  = 0, \quad [ \at \, , \btt ]_{q} =0 \, ,
\ed
\be
[ \tilde{N}_a , \tilde{N}_b ] =0 \, .
\ee

\noindent These considerations can be generalized into the multi-mode case. Considering the $n$
independent bosonic harmonic oscillators $ ( i \ne j )$
\bd
[a_i , a_i^{\dagger} ] = 1, \quad i=1, 2, 3, \ldots,  n \, ,
\ed
\be
[a_i , a_j ] = [ a_i , a_j^{\dagger} ] =0 \,  ,
\ee
leads to the consideration of the following map
\be
\ai = \la_{i-1} a_i  , \quad \ait = a_i^{\dagger}\la_{i-1}^{-1}  \, ,
\ee
where $\la_{i-1}$ is a unitary q-parity operator defined as
\be
\la_{i-1} = \prod_{k=1}^{i-1} \eta_k = \prod_{k=1}^{i-1} q^{ N_k }  
\ee
and
\be
N_k = a_k^{\dagger} a_k  \, .
\ee
Then, one has the commutation relations
\bd
[ \ai , \ait ] =1 , \quad  i =1, 2 , 3, \ldots , n  \,,
\ed
\bd
[ \ai, \aj]_q = 0, \quad  i< j  \, , 
\ed
\bd
[ \ai, \aj]_{\qi} = 0, \quad  i> j \, ,
\ed
\bd
[ \ai, \ajt]_{\qi} = 0, \quad i< j \,
\ed
\bd
[ \ai, \ajt ]_{q} = 0, \quad  i> j  \, ,
\ed
\be
[ \tilde{N}_i , \tilde{N}_j ] =0 \, ,
\ee
where
\be
\tilde{N}_i = \ait \ai \, .
\ee
The second, third, forth and fifth relations can be also written as 
\be
[ \ai, \aj]_{q^{\epsilon_{ij} } } = 0, \quad
[ \ai, \ajt]_{q^{-\epsilon_{ij} } }= 0 \, ,
\ee
where
\be
\epsilon_{ij} =
\begin{cases}
1 \, , \quad  i < j  \cr 0 \, , \quad i = j \cr -1 \,  , \quad  i > j
\end{cases} \, .
\ee

\section{Changing the mutual commutation relations of two different Dirac fields}
For the sake of generality, one may also have a look at the situation where fermion fields are
involved. However, only a simple case involving free fields shall be discussed for the sake of brevity.
Discussing free fields only is not a major disadvantage, since 
a rigorous construction of a non-trivial quntum field theory in four space-time dimensions
has not been successful so far, and most of our practical knowledge in local quantum field theory
is based on considerations concerning free fields acting as operator valued distributions
on a Fock space.\\

\noindent A Dirac field describing non-interacting spin-$\frac{1}{2}$ fermions
like, e.g., the free electron-positron field, can be written in the following form, using
natural units $\hbar=c=1$ and a
relativistic notation with $kx=k_\mu x^\mu= k^0 x^0 -\vec{k} \vec{x}$,
$k^0=\sqrt{\vec{k}^2+m_e^2}>0$ 
\begin{equation}
\psi (x) = \int \frac{d^3 k}{2 k_0 (2 \pi)^3} \sum \limits_{s=\pm \frac{1}{2}}
\{ e^{-ikx} u_s(\vec{k}) a_s(\vec{k}) + e^{+ikx} v_s(\vec{k}) b_s^{\, \dagger}(\vec{k}) \} \, ,
\end{equation}
where the $u_s(\vec{k})$ [$v_s(\vec{k})$] denote electron [positron] spinors for the corresponding
spin $s=\pm \frac{1}{2}$ and momentum $\vec{k} =(k^1,k^2,k^3)$. The electron [positron]
destruction operators $a_s(\vec{k})$ [$b_s(\vec{k})$] and the Hermitian adjoint creation operators
$a_s^{\dagger} (\vec{k})$ [$b_s^{\dagger} (\vec{k})$] fulfill the anti-commutation relations
\begin{displaymath}
\{a_s(\vec{k}),a^{\dagger}_{s'} (\vec{k}') \} = \{b_s(\vec{k}),b^{\dagger}_{s'} (\vec{k}') \} =
\delta_{ss'} 2 k^0 (2 \pi)^3 \delta^{(3)}(\vec{k}-\vec{k}') \, ,
\end{displaymath}
\begin{displaymath}
\{a_s(\vec{k}),a_{s'} (\vec{k}') \} =
\{a^{\dagger}_s(\vec{k}),a^{\dagger}_{s'} (\vec{k}') \} =
\{b_s(\vec{k}),b_{s'} (\vec{k}') \} =
\{b^{\dagger}_s(\vec{k}),b^{\dagger}_{s'} (\vec{k}') \} = 0 \, , 
\end{displaymath}
\begin{equation}
\{a_s(\vec{k}),b_{s'} (\vec{k}') \} =
\{a^{\dagger}_s(\vec{k}),b^{\dagger}_{s'} (\vec{k}') \} =
\{a_s(\vec{k}),b^{\dagger}_{s'} (\vec{k}') \} =
\{a^{\dagger}_s(\vec{k}),b_{s'} (\vec{k}') \} = 0 \, . \label{anticommutation_relations}
\end{equation}
Note that all the destruction operators annihilate the vacuum according to
\begin{equation}
a_s (\vec{k}) |0 \rangle = b_s (\vec{k}) |0 \rangle = 0
\end{equation}
in order to have a Fock Hilbert space representation of free electron and positron states.
Considering an extended theory including a further type of kinematically independent
fermions of mass $m'$, e.g.muons,
one introduces the additional `primed' Dirac field
\begin{equation}
\psi'(x) = \int \frac{d^3 k}{2 k_0' (2 \pi)^3} \sum \limits_{s=\pm \frac{1}{2}}
\{ e^{-ikx} u_s'(\vec{k}) a_s'(\vec{k}) + e^{+ikx} v_s' (\vec{k}) {b_s'}^{\dagger}(\vec{k}) \} \, .
\end{equation}
with $k'_0=\sqrt{\vec{k}^2+m'^2}$ and creation and destruction operators 
fulfilling completely analogous anti-commutation relations as given by eqns.
(\ref{anticommutation_relations}).\\

\noindent It is common practice to assume that the creation and destruction operators
for different fermion types which are (kinematically) independent in the sense that
they cannot be transformed by a Poicar\'e transformations or C, P, or T into each other
\cite{Wigner}
\emph{anticommute}, i.e. one has
\begin{equation}
\{\hat{c},\hat{c}'\}=0 \, , \label{anticommutation}
\end{equation}
where $\hat{c}$ represents any creation or destruction operator for particles of
mass $m$ appearing in eqns. (\ref{anticommutation_relations}) and $\hat{c}'$ a
corresponding creation or destruction operator for particles of mass $m'$.\\

\noindent Introducing the total particle number operator
\begin{equation}
N=N_a + N_{b} + N_{a'} +N_{b'} = \sum \limits_{s=\pm \frac{1}{2}} 
\int \frac{d^3 k}{2 k^0 (2 \pi)^3} \Bigl[ a_s^\dagger (\vec{k}) a_s(\vec{k})+
b_s^\dagger (\vec{k}) b_s(\vec{k})+ {a_s'}^\dagger (\vec{k}) {a_s'}(\vec{k})+
{b_s'}^\dagger (\vec{k}) {b_s'} (\vec{k}) \Bigr]
\end{equation}
one may define the operator
\begin{equation}
\eta = (-1)^N = e^{i \pi N} \, .
\end{equation}
$\eta$ is defined on the whole Fock Hilbert space and fulfills
\begin{equation}
\eta=\eta^\dagger \, , \quad \eta^2=1 \, , \quad \eta^{-1} = \eta^\dagger \, .
\end{equation}
Note that $\eta$ could also be defined by the help of the charge operator
\begin{equation}
\eta=e^{i \pi Q} \, , \quad Q = -N_a + N_{b} - N_{a'} +N_{b'} \, .
\end{equation}
Contrary to the particle number operator $N$, $Q$ is conserved when interactions
are involved and therefore provides some advantages when one tries to discuss
an interacting theory, where the charge structure survives rather than the particle picture.
Since creation and destruction operators change the particle number by $\pm1$, one has
\begin{equation}
\hat{c} \eta=-\eta \hat{c} \, , \, \, \, \mbox{or} \, \, \, \{\eta,\hat{c} \} = 0 \, ,
\end{equation}
and $ \{\eta,\hat{c}' \} = 0$.
Now one may define a new algebra of creation and destruction operators
for particles of mass $m'$, explicitly
\begin{equation}
a'_s(\vec{k}) \mapsto \tilde{a}_s (\vec{k}) = \eta a'_s (\vec{k}) \, , \quad
b'_s(\vec{k}) \mapsto \tilde{b}_s (\vec{k}) = \eta b'_s (\vec{k}) \, ,
\end{equation}
implying
\begin{equation}
{a'_s}^\dagger(\vec{k}) \mapsto {\tilde{a}_s}^\dagger (\vec{k}) = {a'_s}^\dagger
 (\vec{k}) \eta \, , \quad
{b'_s}^\dagger(\vec{k}) \mapsto {\tilde{b}_s}^\dagger (\vec{k}) = {b'_s}^\dagger
(\vec{k}) \eta
\end{equation}
for all $s$ and $\vec{k}$.
This would change the anticommutation relations eq. (\ref{anticommutation}) into commutation
relations
\begin{equation}
[\hat{c}, \tilde{c}]=0 \, ,
\end{equation}
where $\tilde{c}$ is any operator of type
${\tilde{a}_s} (\vec{k})$,
${\tilde{a}_s}^\dagger (\vec{k})$,
${\tilde{b}_s} (\vec{k})$, or
${\tilde{b}_s}^\dagger (\vec{k})$.\\

\end{document}